%% file: Main.tex
\documentclass[conference]{IEEEtran}

\IEEEoverridecommandlockouts
\usepackage{graphicx}
\usepackage{amsmath}
\usepackage{amsfonts}
\usepackage{amssymb}
\usepackage{subfigure}
\usepackage{psfrag}
\usepackage{color}
\usepackage{multicol}
\usepackage{tikz}
\usepackage{verbatimbox}
\usepackage{graphicx}
\usepackage{epsfig}
\usepackage{multirow}
\usepackage{lettrine}
\usepackage{cite}

\usepackage{lipsum,adjustbox}
\usetikzlibrary{calc,positioning,automata}

\newcommand{\be}{\begin{equation}}
\newcommand{\ee}{\end{equation}}

\title{A Generalized Framework for Pulse-Shaping on Delay-Doppler Plane}
\vspace{-1cm}
\author{\IEEEauthorblockN{Mohsen Bayat and Arman Farhang} 
\IEEEauthorblockA{Department of Electronic \& Electrical Engineering, Trinity College Dublin, Ireland \\
\{bayatm, arman.farhang\}@tcd.ie}
\vspace{-0.5cm}
\thanks{This publication has emanated from research conducted with the financial support of Science Foundation Ireland under Grant numbers SFI/19/FFP/7005(T) and SFI/21/US/3757. For the purpose of Open Access, the authors have applied a CC BY public copyright licence to any Author Accepted Manuscript version arising from this submission.}
}

\begin{document}
\maketitle
\begin{abstract}
The primary objective of this paper is to establish a generalized framework for pulse-shaping on the delay-Doppler plane. To this end, we classify delay-Doppler pulse-shaping techniques into two types, namely, circular and linear pulse-shaping. This paves the way towards the development of a generalized pulse-shaping framework. Our generalized framework provides the opportunity to compare different pulse-shaping techniques under the same umbrella while bringing new insights into their properties. In particular, our derivations based on this framework reveal that the recently emerged waveform orthogonal delay-Doppler multiplexing modulation (ODDM) is a linear pulse-shaping technique. By presenting ODDM under our generalized framework, we clearly explain the observed staircase behavior of its spectrum which has not been previously reported in the literature. Another contribution of this paper is proposal of a simple out-of-band (OOB) emission reduction technique by inserting a small number of zero-guard (ZG) symbols along the delay dimension of the circularly pulse-shaped signals. Additionally, inserting the zero-guards improves the bit-error-rate (BER) performance of both circular and linear pulse-shaping techniques. Finally, our simulation results confirm the validity of our mathematical derivations, claims and the effectiveness of the ZGs in OOB reduction and BER performance improvement.
\end{abstract}
\vspace{-0.2cm}
\input{Introduction}
\input{Principle}
\vspace{-0.1cm}
\input{CPS}

\vspace{-0.1cm}
\input{LPS}
\vspace{-0.2cm}
\input{GFW}
\vspace{0cm}
\input{BER_Results}
\vspace{0cm}
\input{Conclusion}
\vspace{0cm}

\bibliographystyle{IEEEtran} 
\bibliography{IEEEabrv,references}

\end{document}

%% file: Introduction.tex
\section{introduction}\label{sec:Introduction}
Emerging applications, use cases, and network architectures such as autonomous driving, high-speed trains, and interconnected aerial and terrestrial systems bring new challenges to future wireless networks \cite{Yang2019}. Time-varying wireless environment is a common aspect of all these systems. While orthogonal frequency division multiplexing (OFDM) has been enjoying its dominance in the last two generations of wireless systems, it cannot cope with the highly time-varying wireless channels that appear in future networks \cite{Hadani2017}.

Initiated by the landmark paper of Hadani et al. \cite{Hadani2017}, multiplexing data in delay-Doppler domain has led to a great deal of interest among researchers in both academia and industry. As a paradigm-shifting technology, delay-Doppler multiplexing transforms the time-varying channel into a two-dimensional (2D) time-invariant one in delay-Doppler domain \cite{Hadani2017}. 
Orthogonal time-frequency space (OTFS), \cite{Hadani2018}, is a rudimentary delay-Doppler multiplexing technique that can be implemented on the top of any multicarrier system, \cite{Hadani2017,Hadani2018,Tiwari2020,Pishva2022,Andrea2023}. This is because the delay-Doppler domain data can be easily translated into the time-frequency domain by a 2D Fourier transform and vice versa. More recently, a new modulation technique called orthogonal delay-Doppler multiplexing (ODDM) was introduced in \cite{Lin2022_1}. ODDM deploys orthogonal pulses in delay-Doppler domain to directly map the data symbols to the delay-time domain \cite{Lin2022_2}. From a different viewpoint, ODDM is a multicarrier system whose signal is formed by staggering multiple pulse-shaped OFDM symbols.

As mentioned earlier, OTFS can be implemented using any time-frequency data transmission technology \cite{Hadani2017}. Backward compatibility with the existing wireless standards is an important aspect that requires careful attention in the air interface design for future networks. This makes OFDM-based delay-Doppler multiplexing techniques quite attractive for deployment in next generation networks. Hence, the focus of this paper is specifically on OFDM-based OTFS and ODDM.

There exist a number of OFDM-based delay-Doppler multiplexing techniques, \cite{Raviteja2019b, Raviteja2018b, Zhou2023, Lin2022_0, Wei2021_2, Tusha2023, Saifkhan2022, Lin2022_1, Lin2022_2, Li2023}.
While the authors in \cite{Raviteja2019b} consider ideal pulse-shaping, the impact of rectangular pulse-shaping on OTFS is investigated in \cite{Raviteja2018b}. More recently, ideal and rectangular pulse-shaping techniques are compared in \cite{Zhou2023}.
The authors in \cite{Lin2022_0} reveal that rectangular pulses in OTFS lead to increased interference at the edges of the delay blocks at the receiver side. 
Beyond studying the pulse-shaping effects on OTFS, the authors in \cite{Wei2021_2} and \cite{Tusha2023} focus on pulse-shape design. The main idea of \cite{Wei2021_2} is to design an optimal receiver window across the time dimension for interference cancellation purposes. To reduce the Doppler-induced leakage at the receiver, a global windowing method was introduced in \cite{Tusha2023}.
In addition to the literature on pulse-shape design, there are also several studies that introduce various pulse-shaping techniques.
The authors in \cite{Saifkhan2022} and \cite{Li2023} demonstrate that the initial two-stage OTFS proposal in \cite{Hadani2017} can be effectively represented by a one-stage Zak transform, including the necessary pulse-shaping for OTFS in the delay-Doppler domain. These two works primarily focus on the feasibility of the existing pulse-shaping filter in the delay-Doppler plane.

Despite the numerous works on this topic, there is a lack of in-depth analysis to consolidate and compare various pulse-shaping techniques within a generalized framework.
Additionally, in the earlier literature on delay-Doppler multiplexing, oversampling has been disregarded due to the focus on critically sampled baseband signals.
Nevertheless, multi-stage digital oversampling is a necessity for practical implementation.

Hence, in this paper, we present the discrete-time formulation of the OFDM-based pulse-shaping techniques on the delay-Doppler plane with oversampling. 
We classify these techniques into two categories of linear and circular pulse-shaping. 
Then, we derive a generalized framework that sheds light on the properties, differences and similarities of various pulse-shaping techniques. 
In this study, we establish a discrete-time representation of ODDM. 
Our derivations reveal that ODDM is a linear pulse-shaping technique. Furthermore, by presenting ODDM within the context of our generalized framework, we clearly explain the observed staircase behavior of its spectrum which we report for the first time in this paper.
To enhance the out-of-band (OOB) emission performance of the circular pulse-shaping techniques, we propose inserting a small number of zero-guard (ZG) symbols along the delay dimension. As mentioned earlier, the channel imposed interference is severe at the edges of the delay blocks. Hence, ZG insertion improves the bit-error-rate (BER) performance. Finally, we numerically confirm our mathematical derivations and claims by simulations. 

The rest of this paper is organized as follows.
Section~\ref{sec:Principle} covers the delay-Doppler signaling fundamentals. Circular and linear pulse-shaping techniques are discussed in Sections~\ref{sec:CPS} and \ref{sec:LPS}. An OOB emission reduction technique for circularly pulse-shaped signals is also proposed in Section~\ref{sec:CPS}. Using the derivations in Sections~\ref{sec:CPS} and \ref{sec:LPS}, a generalized framework for delay-Doppler plane pulse-shaping is introduced in Section~\ref{sec:FW}. Finally, our numerical results are presented in Section~\ref{sec:Numerical} and the paper is concluded in Section~\ref{sec:Conclusion}.

\textit{Notations}: $L$-fold upsampling and downsampling along a given dimension, $l$, of a multidimensional signal are represented by the operators $( \cdot )_{l,\uparrow_{L}}$ and $( \cdot )_{l,\downarrow_{L}}$, respectively. $M$-point circular convolution is dented as $\stackrel{\mbox{\tiny{$M$}}}{\circledast}$ and $*$ represents linear convolution. $\delta[\cdot]$ is the Dirac delta function and $(\cdot)^{-1}$ indicates the inverse operation. The discrete Fourier transform (DFT) of $x[l,n]$ with respect to $n$ is represented as $X[l,k] = \mathcal{F}_{N,n} \{ x[l,n] \} \!=\!\! \frac{1}{\sqrt{N}}\sum_{n=0}^{N-1} x[l,n] e^{-j\frac{2\pi kn}{N}}$. 

%% file: Principle.tex
\section{System Model}\label{sec:Principle}
\vspace{-0.1cm}
We consider the quadrature amplitude modulated (QAM) data symbols, $D[l,k]$, on a regular grid in the delay-Doppler domain, where $l=0,\ldots,M-1$ and $k=0,\ldots,N-1$ represent the delay and Doppler indices, respectively. Given the delay spacing of $\Delta \tau$, each delay block comprising $M$ delay bins has the duration of $T=M\Delta \tau$. Hence, the Doppler spacing is $\Delta \nu = \frac{1}{NT}$.
To form the transmit signal, the data symbols are translated to the delay-time domain by an inverse DFT (IDFT) operation along the Doppler dimension as
\begin{align} \label{eqn:ifft} X[l,n] = \frac{1}{\sqrt{N}} \sum_{k=0}^{N-1} D[l,k] e^{j \frac{2 \pi kn}{N}}. \end{align}
Then, the samples on the delay-time grid, $X[l,n]$ are converted to the serial stream $x[\kappa]\!\!=\!\!X[l,n]$ for $\kappa\!=\!nM\!+l$, $n\!=\!0,\!\ldots\!,N\!-\!1$, and $l=0,\ldots,M-1$ and a CP is appended at the beginning of the block. The CP length is chosen as $L_{\rm{cp}} \geq L_{\rm{ch}}$ to prevent inter-block interference, where $L_{\rm{ch}}$ is the channel length.

Let us consider the received signal after transmission over the linear time-varying (LTV) channel and CP removal as $r[\kappa]$ for $\kappa=0,\ldots,MN-1$. To obtain the received data symbols in the delay-Doppler domain that are affected by the channel, we first form the 2D delay-time signal $Y[l,n]=r[nM+l]$, where $l=0,\ldots, M-1$ and $n=0,\ldots, N-1$ are delay and time indices, respectively.
Then, we perform an $N$-point DFT operation on $Y[l,n]$ along the time dimension to obtain
\be \label{eqn:D_hat} \widetilde{D}[l,k] = \frac{1}{\sqrt{N}}\sum_{n=0}^{N-1} Y[l,n] e^{-j \frac{2 \pi kn}{N}}. \ee
Finally, the received symbols $\widetilde{D}[l,k]$ are passed through the channel equalizer to estimate the transmitted data symbols.

The above formulation is generic to delay-Doppler multiplexing techniques and it does not capture the pulse-shaping effects. Pulse-shaping is a design consideration that is of a paramount importance to any other communication system.
Hence, in the rest of this paper, we first classify different pulse-shaping techniques into two types, namely, circular and linear pulse-shaping. Then, we introduce a generalized framework that reveal the properties of different pulse-shaping techniques. 

%% file: CPS.tex
\section{Circular Pulse-shaping}\label{sec:CPS}

\begin{figure*}
  \centering 
  {\includegraphics[scale=0.084]{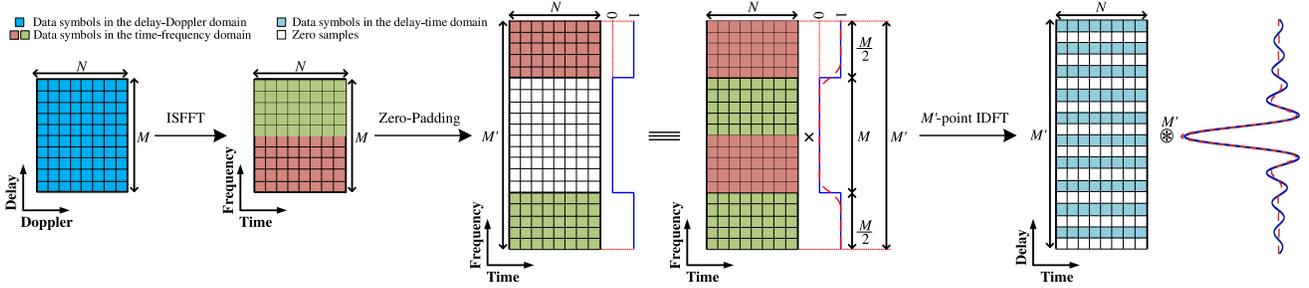}}
  \vspace{-0.2cm}
  \caption{The relationship between zero-padding in frequency dimension and oversampling in delay dimension where $M=10$, $N=8$, $L_{\rm{us}}=2$.} 
  \vspace{-0.2cm}
  \label{fig:zp}
\end{figure*}

In this section, we initiate our discussion with oversampling and pulse-shaping the multiplexed data in delay-Doppler domain, using the OFDM modulator. For reasons that will become clear shortly, this type of pulse-shaping falls under the category of circular pulse-shaping techniques.

The initial proposal of OTFS utilizes the inverse symplectic finite Fourier transform (ISFFT) to convert data symbols into the time-frequency domain. This is followed by an OFDM modulator, which performs oversampling and pulse-shaping, by zero-padding in the frequency domain and windowing in the time domain, often with a rectangular window. With the oversampling factor $L_{\rm{us}}$, this process is performed by increasing the number of frequency bins from $M$ to $M'=ML_{\rm{us}}$ and setting the high-frequency bins to zero \cite{Farhang2010}. As shown in Fig.~\ref{fig:zp}, this can be alternatively performed by creating $L_{\rm{us}}$ replicas of the signal along the frequency dimension and multiplication by a rectangular window in each time-slot. 

Creating $L_{\rm{us}}$ spectral replicas of the time-frequency domain signal at each time-slot is equivalent to the expansion of the delay-time domain signal in (\ref{eqn:ifft}) along the delay dimension as
\be \label{eqn:us}  X_{\rm{e}}[l',n] = \sum_{l=0}^{M-1}X[l,n] \delta[l'-lL_{\rm{us}}], \ee
where $l'=0,\ldots,M'-1$. 
Frequency domain multiplication by a rectangular window  can also be equivalently performed in the delay-time domain with an $M'$-point circular convolution operation along the delay dimension, i.e., 
\begin{align} \label{eqn:cotfs_ps} \overline{X}^{\mathtt{C}}[l',n] = X_{\rm{e}}[l',n] \stackrel{\mbox{\tiny $~~M'$}}{\circledast} p[l'], \end{align} 
where $p[l']$ is the `sinc' function which can be replaced with any Nyquist pulse such as the root-raised cosine (RRC) pulse.

The transmit signal before CP addition can be constructed by converting the two-dimensional delay-time domain signal, $\overline{X}^{\mathtt{C}}[l',n]$, into the serial stream $x_{\rm us}^{\mathtt{C}}[\kappa']=\overline{X}^{\mathtt{C}}[l',n]$ where $\kappa'=nM'+l'$.
Assuming $\overline{X}^{\mathtt{C}}[l',n]$ to be non-zero only for $l'=0,\ldots, M'-1$ and $n=0,\ldots, N-1$, and zero elsewhere, $x_{\rm us}^{\mathtt{C}}[\kappa']$ can be expressed as
\begin{align} \label{eqn:cotfs_p/s} x_{\rm us}^{\mathtt{C}}[\kappa'] = \sum_{n=0}^{N-1} \overline{X}^{\mathtt{C}}[\kappa'-nM',n]. \end{align}
By substituting (\ref{eqn:cotfs_ps}) into (\ref{eqn:cotfs_p/s}), we obtain
\begin{align} \label{eqn:cotfs_p/s2} x_{\rm us}^{\mathtt{C}}[\kappa'] &= \sum_{n=0}^{N-1} X_{\rm{e}}[\kappa'-nM',n] \stackrel{\mbox{\tiny{$~~M'$}}}{\circledast} p[\kappa'-nM']. \end{align}
As shown below, using (\ref{eqn:ifft}) in (\ref{eqn:us}) and reordering the summations, $X_{\rm{e}}[l',n]$ can be directly obtained by expanding the delay-Doppler domain data symbols.
\begin{align} \label{eqn:cotfs_p/s3}
X_{\rm{e}}[l',n] &\!=\! \sum_{l=0}^{M\!-\!1} \! \left( \frac{1}{\sqrt{N}} \sum_{k=0}^{N\!-\!1} D[l,k] e^{j \frac{2 \pi nk}{N}} \right) \delta[l'-lL_{\rm{us}}] \nonumber \\ &= \frac{1}{\sqrt{N}} \sum_{k=0}^{N-1} \left( \sum_{l=0}^{M-1} D[l,k] \delta[l'-lL_{\rm{us}}] \right) e^{j \frac{2 \pi nk}{N}} \nonumber \\ &=\! \frac{1}{\sqrt{N}}\! \sum_{k=0}^{N-1} D_{\rm{e}}[l',k] e^{j \frac{2 \pi nk}{N}} \!= \mathcal{F}_{N,k}^{-1} \Bigl\{ D_{\rm{e}}[l',k] \Bigr\}, \end{align}
where $D_{\rm{e}}[l',k]=\Big( D[l,k] \Big)_{l,\uparrow_{L_{\rm us}}}$. 
Then, by inserting (\ref{eqn:cotfs_p/s3}) into (\ref{eqn:cotfs_p/s2}), we obtain
\begin{align} \label{eqn:cotfs_p/s4} x_{\rm{us}}^{\mathtt{C}}[\kappa'] &= \!\! \sum_{n=0}^{N-1} \mathcal{F}_{N,k}^{-1} \Bigl\{ D_{\rm{e}}[\kappa'\!-\!nM',k] \Bigr\} \stackrel{\mbox{\tiny{$~~M'$}}}{\circledast} p[\kappa'\!-\!nM'] \nonumber\\ &= \!\! \sum_{n=0}^{N-1} \! \mathcal{F}_{N,k}^{-1} \Bigl\{ D_{\rm{e}}[\kappa'\!-\!nM',k] \stackrel{\mbox{\tiny{$~~M'$}}}{\circledast} p[\kappa'\!-\!nM'] \Bigr\}. \end{align}
Our result in (\ref{eqn:cotfs_p/s4}) reveals that oversampling and pulse-shaping can be performed in either the delay-Doppler or delay-time domains.
In fact, circular convolution in (\ref{eqn:cotfs_p/s4}) can be interpreted as making $D_{\rm{e}}[l',k]$ and $p[l']$ periodic along the delay dimension, linearly convolving them, and confining the resulting signal by a rectangular window with length $M'$. This can be thought of as a circular pulse-shaping procedure for OTFS which we call C-PS OTFS. After appending a CP with length $L'_{\rm{cp}}=L_{\rm{cp}}L_{\rm{us}}$ at the beginning of $x_{\rm{us}}^{\mathtt{C}}[\kappa']$, the resulting signal is transmitted through the channel.

At the receiver, we express the received signal after CP removal as $r_{\rm{us}}^{\mathtt{C}}[\kappa']$. 
Then, the 2D delay-time signal $Y_{\rm{us}}^{\mathtt{C}}[l',n]=r_{\rm{us}}^{\mathtt{C}}[l'+nM']$ is constructed where $l'=0,\ldots,M'-1$ and $n=0,\ldots,N-1$.
Next, $Y_{\rm{us}}^{\mathtt{C}}[l',n]$ goes through a matched filtering process using $p^*[-l']$, followed by down-sampling with a factor of $L_{\rm ds}\!=\!L_{\rm{us}}$, both along the delay dimension. Finally, we take an $N$-point DFT across the time dimension to obtain the received delay-Doppler domain data symbols as
\begin{align} \label{eqn:D_tilde} \widetilde{D}[l,k] &= \mathcal{F}_{N,n} \Bigl\{ \Big(Y_{\rm{us}}^{\mathtt{C}}[l',n] \stackrel{\mbox{\tiny{$~~M'$}}}{\circledast} p[l']\Big)_{\downarrow_{L_{\rm ds}}}\Bigr\} \nonumber\\ &=  \Big(\mathcal{F}_{N,n} \Bigl\{  Y_{\rm{us}}^{\mathtt{C}}[l',n] \Bigr\} \stackrel{\mbox{\tiny{$~~M'$}}}{\circledast} p[l']\Big)_{\downarrow_{L_{\rm ds}}}, \end{align}
where $p^*[-l']=p[l']$ as $p[l']$ is symmetric and real-valued.
As shown in (\ref{eqn:D_tilde}), matched filtering and down-sampling stages can be performed in either delay-Doppler or delay-time domain.

The circular convolution in (\ref{eqn:cotfs_ps}) causes the transients of the pulse-shape to wrap around at the edges of each delay block. 
Therefore, circular pulse-shaping does not allow the transients of the
pulse shape to appear. This leads to abrupt changes at the
boundaries of the delay blocks that result in high OOB emissions.
To circumvent the wrap around effect, we propose zero-padding the delay-Doppler domain data symbols in all time-slots along the delay dimension. Inserting ZG symbols lead to the appearance of the smooth transients at the edges of each delay block which lowers the OOB emissions.
As we will show in Section~\ref{sec:Numerical}, by inserting only a small number of ZG symbols, a significant amount of OOB reduction can be achieved. 
Furthermore, the benefits of the ZG insertion is two-fold as they also lead to BER performance improvement.

%% file: LPS.tex
\section{Linear Pulse-shaping}\label{sec:LPS}
In the previous section, we discussed and formulated circular pulse-shaping while explaining OOB emission limitations of this pulse-shaping class. In this section, we explain and formulate linear pulse-shaping as an alternative to circular pulse-shaping which inherently has low OOB emissions.
Additionally, we present the discrete-time formulation of the emerging delay-Doppler multiplexing technique, ODDM. Our derivations show that ODDM is a linear pulse-shaping technique while explaining an interesting property of this waveform that has not been previously reported in the literature.

\subsection{Linearly Pulse-Shaped OTFS} \label{subsec:LPS-OTFS}
As explained in Section~\ref{sec:CPS}, frequency domain oversampling is performed in two steps. At the first step, the translated data symbols to the time-frequency domain are repeated $L_{\rm us}$ times along frequency. At the second step, only one replica is preserved. In circular pulse-shaping, the second step is performed by cyclic filtering, see (\ref{eqn:cotfs_ps}). In this section, we alternatively remove the unwanted spectral replicas by linear filtering. Linear filtering leads to smooth signal edges in each time-slot and thus, low OOB emissions.
For a specific time-slot $n$, linear pulse-shaping can be performed by replacing circular convolution in (\ref{eqn:cotfs_ps}) with linear convolution, 
\begin{align} \label{eqn:lotfs_ps} \overline{X}^{\mathtt{L}}[l',n] = X_{\rm{e}}[l',n] * p[l'], \end{align}
where $l'\!\!=\!-\frac{M'}{2}+1,\!\ldots\!, M'+\frac{M'}{2}-1$. 
Thanks to the Nyquist property of the deployed pulses for pulse-shaping, the transmit signal before adding the CP can be constructed by overlapping the adjacent time slots with a spacing of $M'$ samples as 
\begin{align} \label{eqn:lotfs_p/s} x^{\mathtt{L}}_{\rm{us}}[\kappa'] &= \sum_{n=0}^{N-1} \overline{X}^{\mathtt{L}}[\kappa'-nM',n], \end{align}
where $\kappa'\!=\!-\frac{M'}{2}+1,\ldots, M'N\!+\!\frac{M'}{2}\!-\!1$.
It is important to note that the first ${M'}/{2}-1$ and the last ${M'}/{2}$ samples of $x_{\rm{us}}^{\mathtt{L}}[\kappa']$ are part of the transient interval of the pulse-shaping filter $p[l']$. To shorten the transient intervals, the pulse $p[l']$ can be truncated up to $Q$ zero-crossings, which include significant sidelobes on each side of its main lobe, \cite{Lin2022_2}. 
However, pulse truncation comes at the expense of approximating the Nyquist criterion, which means perfect reconstruction of the transmit symbols at the receiver cannot be assured, \cite{Farhang2010}. While pulse truncation does not have a substantial impact on the BER performance for small constellation sizes, it can result in a performance penalty as the constellation size increases.
Moreover, for very small values of $Q$, truncation results in significantly increased OOB emissions. Thus, $Q$ should be chosen carefully.

Using (\ref{eqn:cotfs_p/s3}) and (\ref{eqn:lotfs_ps}), (\ref{eqn:lotfs_p/s}) can be rearranged as
\begin{align} \label{eqn:lotfs_p/s3} x_{\rm us}^{\mathtt{L}}[\kappa'] &=\sum_{n=0}^{N-1} X_{\rm{e}}[\kappa'-nM',n] * p[\kappa'-nM'] \nonumber\\ &= \sum_{n=0}^{N-1} \mathcal{F}_{N,k}^{-1} \Bigl\{ D_{\rm{e}}[\kappa'\!-\!nM',k] \Bigr\} * p[\kappa'\!-\!nM'] \nonumber\\ &= \sum_{n=0}^{N-1} \! \mathcal{F}_{N,k}^{-1} \Bigl\{ D_{\rm{e}}[\kappa'\!-\!nM',k] * p[\kappa'\!-\!nM'] \Bigr\}. \end{align}
Considering the truncated pulse $p[l']$, $\kappa'\!=\!-Q',\ldots, M'N+Q'-1$, and $Q'=QL_{\rm{us}}$ for $Q\in\{1,\ldots,\frac{M}{2}\}$.
From (\ref{eqn:lotfs_p/s3}), it is evident that oversampling and pulse-shaping can be performed either in the delay-Doppler domain or the delay-time domain.
This flexibility is due to the fact that pulse-shaping is carried out along the delay dimension and independent of Doppler. A CP of length $L'_{\rm{cp}}$ is then inserted at the beginning of $x_{\rm us}^{\mathtt{L}}[\kappa']$. As this type of pulse-shaping is based on OTFS framework, we refer to it as linearly pulse-shaped OTFS (L-PS OTFS).
 
Similarly to C-PS OTFS in Section~\ref{sec:CPS}, we can represent the received signal after CP removal as $r_{\rm{us}}^{\mathtt{L}}[\kappa']$. Then, we can rearrange the received signal samples to form the 2D signal $Y_{\rm{us}}^{\mathtt{L}}[l',n]=r_{\rm{us}}^{\mathtt{L}}[l'+nM']$ for $l'=-Q',\ldots,M'+Q'-1$ and $n=0,\ldots,N-1$ in the delay-time domain.
The received delay-Doppler domain symbols can be obtained through a two-step procedure, i.e., matched filtering followed by down-sampling along the delay dimension with a factor of $L_{\rm{ds}}=L_{\rm{us}}$.
This operation can also be performed either in the delay-time domain or the delay-Doppler domain as
\begin{align} \label{eqn:Dl_tilde} \widetilde{D}[l,k] &= \mathcal{F}_{N,n} \Bigl\{ \Big(Y_{\rm{us}}^{\mathtt{L}}[l',n] * p[l']\Big)_{\downarrow_{L_{\rm ds}}}\Bigr\} \nonumber\\ &=  \Big(\mathcal{F}_{N,n} \Bigl\{  Y_{\rm{us}}^{\mathtt{L}}[l',n] \Bigr\} * p[l']\Big)_{\downarrow_{L_{\rm ds}}}, \end{align}
where $p^*[-l']=p[l']$, since it is real and symmetric.

\subsection{ODDM as a Linear Pulse-Shaping Technique}\label{subsection:oddm}

\begin{figure*}
\centering
\begin{adjustbox}{width=\textwidth}
\tikzstyle{block} = [rectangle, fill=white, minimum width=0.5cm, minimum height=0.8cm, text centered, draw=black, align=center]
\tikzstyle{arrow} = [thick,->,>=stealth]

\begin {tikzpicture}[-latex, auto, node distance=1.9cm, on grid,semithick,
state/.style ={circle, top color=white, bottom color= processblue!20, draw, processblue, minimum width =0.5 cm}]

\node (pro1) {$D[l,k]$}; 
\node (pro2) [block, right of=pro1] {$\big( \cdot \big)_{l,\uparrow_{L_{\rm us}}}$};
\node (pro3) [block, right of=pro2] {$p_{k}[l']$};
\node (pro4) [block, right of=pro3] {$\mathcal{F}^{-1}_{N,k}\{\cdot\}$};
\node (pro5) [block, right of=pro4] {O/A};
\node (pro6) [block, right of=pro5] { CP \\ Addition };
\node (pro7) [block, right of=pro6] { LTV \\ Channel };
\node (pro8) [block, right of=pro7] { CP \\ Removal};
\node (pro9) [block, right of=pro8] {S/P};
\node (pro10) [block, right of=pro9] {$\mathcal{F}_{N,k}\{\cdot\}$};
\node (pro11) [block, right of=pro10] {$p_{k}^*[-l']$};
\node (pro12) [block, right of=pro11] {$\big( \cdot \big)_{l',\downarrow_{L_{\rm ds}}}$};
\node (pro13) [right of=pro12] {$\widetilde{D}[l,k]$};

\draw [arrow] (pro1) -- (pro2); \draw [arrow] (pro2) -- (pro3); \draw [arrow] (pro3) -- (pro4); \draw [arrow] (pro4) -- (pro5); \draw [arrow] (pro5) -- (pro6); \draw [arrow] (pro6) -- (pro7); \draw [arrow] (pro7) -- (pro8); \draw [arrow] (pro8) -- (pro9); \draw [arrow] (pro9) -- (pro10); \draw [arrow] (pro10) -- (pro11); \draw [arrow] (pro11) -- (pro12); \draw [arrow] (pro12) -- (pro13);

\draw [color=gray,thick,dashed](0.7,-0.8) rectangle (10.4,0.8);
\node at (0,1.1) [above=30mm, right=49mm] {{Modulator}};

\draw [color=gray,thick,dashed](12.3,-0.8) rectangle (22,0.8);
\node at (10,1.1) [above=30mm, right=60mm] {{Demodulator}};

\end{tikzpicture}
\end{adjustbox}
\caption{Generalized modem structure for pulse-shaping on delay-Doppler plane (for C/L-PS OTFS, $p_k[l']=p_0[l'],~\forall k$).}
\vspace{-0.1cm}
\label{fig:imp}
\end{figure*}
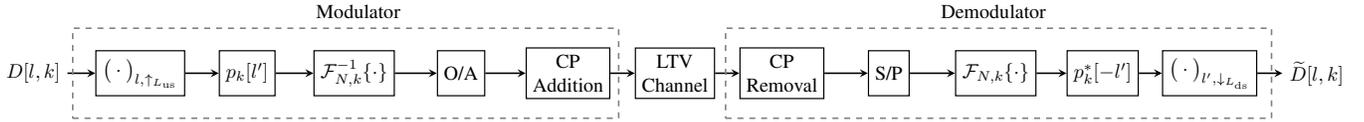

In ODDM, data transmission relies on an orthogonal delay-Doppler domain pulse. To construct the transmit signal, each data symbol $D[l,k]$, scales the pulse-shape that is shifted by $l$ positions along the delay dimension and modulated to the Doppler frequency $k/NT$, i.e.,
\be \label{eqn:Cont_DT} x^{\mathtt{ODDM}}(t) = \sum_{l=0}^{M-1} \sum_{k=0}^{N-1} D[l,k] u(t-l\frac{T}{M}) e^{j \frac{2 \pi k}{NT}(t-l\frac{T}{M})}, \ee
where $-\frac{T}{2} \!\! \leq \!\! t \!\! < \!\! NT+\frac{T}{2}$, $T\!\!=\!\!M \Delta \tau$ is the symbol period, $u(t)=\sum_{n=0}^{N-1} p(t-nT)$, and $p(t)$ is a Nyquist pulse that may be truncated.
After appending a CP with length $L_{\rm{cp}}T_{\rm{s}}$ at the beginning of $x^{\mathtt{ODDM}}(t)$, the resulting signal $x^{\mathtt{ODDM}}_{\rm{cp}}(t)$ is transmitted over the channel.

At the receiver, the CP is discarded first and then the received symbols are obtained after matched filtering, \cite{Lin2022_2}. In the following, we proceed to establish the discrete-time formulation of ODDM that sheds light on some of its unexplored aspects.
Considering the oversampling factor $L_{\rm{us}}$, (\ref{eqn:Cont_DT}) can be represented in discrete time as
\begin{align} \label{eqn:Disc_DT1}
x_{\rm{us}}^{\mathtt{ODDM}}[\kappa'] = &\sum_{l=0}^{M-1} \sum_{k=0}^{N-1} D[l,k] u[\kappa'-lL_{\rm{us}}] e^{j \frac{2\pi k(\kappa'-lL_{\rm{us}})}{NM'}},
\end{align}
where $u[\kappa']=\sum_{n=0}^{N-1}p[\kappa'-nM']$ for $\kappa'=-Q',\ldots,M'N+Q'-1$ and $x_{\rm{us}}^{\mathtt{ODDM}}[\kappa'] = x^{\mathtt{ODDM}}(\kappa' \frac{T}{M'})$, $u[\kappa'-lL_{\rm{us}}]=u((\kappa' -lL_{\rm{us}})\frac{T}{M'})$.
Hence, (\ref{eqn:Disc_DT1}) can be expanded as
\begin{align} \label{eqn:Disc_DT2}
x_{\rm{us}}^{\mathtt{ODDM}}[\kappa'] \!&= \!\sum_{n=0}^{N\!-\!1} \sum_{l=0}^{M\!-\!1} \sum_{k=0}^{N\!-\!1} D[l,k] p[\kappa'\!-\!lL_{\rm{us}}\!-\!nM'] e^{j \frac{2\pi k(\kappa'-lL_{\rm{us}})}{NM'}}
\nonumber\\
&= \!\sum_{n=0}^{N\!-\!1} \sum_{k=0}^{N\!-\!1} \! \left(\sum_{l=0}^{M\!-\!1} \! D[l,k] p_k[(\kappa'\!-\!nM')\!-\!lL_{\rm{us}}] \right) \! e^{j \frac{2 \pi kn}{N}} 
\nonumber\\
&= \!\!\sum_{n=0}^{N\!-\!1} \sum_{k=0}^{N\!-\!1} \! \left(\sum_{\lambda=0}^{M'\!-\!1} \!\! D[\frac{\lambda}{L_{\rm{us}}},k] p_k[(\kappa'\!-\!nM')\!-\!\!\lambda] \right) \! e^{j \frac{2 \pi kn}{N}} 
\nonumber\\
&=\!\! \sum_{n=0}^{N\!-\!1} \sum_{k=0}^{N\!-\!1} \left( D[\frac{\kappa'-nM'}{L_{\rm{us}}},k] * p_k[\kappa'\!-nM'] \right) e^{j \frac{2\pi kn}{N}}
\nonumber\\
&= \!\sqrt{N} \sum_{n=0}^{N-1} \mathcal{F}_{N,k}^{-1} \Bigl\{ D_{\rm{e}}[\kappa'-nM',k] * p_k[\kappa'\!-\!nM'] \Bigr\},
\end{align}
where the second line in (\ref{eqn:Disc_DT2}) is obtained by multiplication of the first line by $e^{-j \frac{2 \pi k nM'}{NM'}} e^{j \frac{2 \pi k nM'}{NM'}}=1$ and defining the modulated pulse $p_k[\kappa']\!=\!p[\kappa']e^{j \frac{2 \pi k \kappa'}{NM'}}$. The third line is obtained by a change of variable $\lambda=lL_{\rm{us}}$.

In its initial proposal, ODDM was described as a combination of $M$ shifted and staggered pulse-shaped OFDM signals, $x_{l}^{\mathtt{ODDM}}(t)=\sum_{k=0}^{N-1}D[l,k] e^{j \frac{2 \pi k}{NT}t} u(t)$, each with $N$ Doppler subcarriers \cite{Lin2022_1,Lin2022_2}. This results in the ODDM transmit signal $x^{\mathtt{ODDM}}(t) = \sum_{l=0}^{M-1} x_l^{\mathtt{ODDM}}(t - l\frac{T}{M})$.
A more recent work presented in \cite{Tong2023} approximated the ODDM transmit signal by sample-wise pulse-shaping of the serialized delay-time signal $X[l,n]$, resulting in the same signal as the one in (\ref{eqn:lotfs_p/s3}). 
However, this approximation is accurate only when $2Q\ll M$.

Comparing our derivation in (\ref{eqn:Disc_DT2}) with (\ref{eqn:lotfs_p/s3}), it becomes evident that ODDM is a linear pulse-shaping technique, and it is different from L-PS OTFS. The distinction between ODDM and L-PS OTFS primarily lies in the choice of pulse-shaping filter for different Doppler bins. In ODDM, pulse-shaping along the delay dimension for a specific Doppler bin $k$ is performed by the modulated pulse $p_k[\kappa']$.
This clearly explains the staircase behavior of the ODDM spectrum that is observed in our numerical results in Section~\ref{sec:Numerical}.
In contrast, L-PS OTFS employs the same pulse $p[\kappa']$, for pulse-shaping along the delay dimension across all Doppler bins, as shown in (\ref{eqn:lotfs_p/s3}).

Considering the received signal after CP removal, $r_{\rm{us}}^{\mathtt{ODDM}}[\kappa']$, the 2D delay-time domain signal can be represented as $Y_{\rm{us}}^{\mathtt{ODDM}}[l',n]=r_{\rm{us}}^\mathtt{ODDM}[l'+nM']$, where $l'=-Q',\cdots,M'+Q'-1$ and $n=0,\ldots,N-1$. 
Due to the dependency of the pulse-shaping filter on the Doppler index, pulse-shaping at the transmitter and matched filtering at the receiver can only be performed in the delay-Doppler domain. This is another point of difference between ODDM and L-PS OTFS. Thus, to obtain the received symbols, we first convert the signal $Y_{\rm{us}}^{\mathtt{ODDM}}[l',n]$ to the delay-Doppler domain. Then, noting that $p^*_k[-l'] \neq p_k[l']$, we apply the matched filter, $p^*_k[-l']$, followed by down-sampling, both along the delay dimension which is mathematically expressed as
\begin{align} \label{eqn:Do_tilde} \widetilde{D}[l,k] &=  \frac{1}{\sqrt{N}} \Big(\mathcal{F}_{N,n} \Bigl\{  Y_{\rm{us}}^{\mathtt{ODDM}}[l',n] \Bigr\} * p^*_k[-l']\Big)_{\downarrow_{L_{\rm ds}}}. \end{align}

%% file: GFW.tex
\section{Generalized Pulse-shaping Framework}\label{sec:FW}
Based on our derivations in Sections~\ref{sec:CPS} and \ref{sec:LPS}, in this section, we introduce a generalized framework that encompasses different pulse-shaping techniques on delay-Doppler plane that were discussed earlier. This framework provides a deep understanding of the properties, similarities, and distinctions between the pulse-shaping techniques under study. Based on the transmit signal expressions in (\ref{eqn:cotfs_p/s4}), (\ref{eqn:lotfs_p/s3}) and (\ref{eqn:Disc_DT2}) and the receive signal expressions in (\ref{eqn:D_tilde}), (\ref{eqn:Dl_tilde}), and (\ref{eqn:Do_tilde}), the generalized baseband block diagram for pulse-shaping on the delay-Doppler plane is illustrated in Fig.~\ref{fig:imp}.

As shown in Fig.~\ref{fig:imp}, all of these pulse-shaping techniques can be implemented in four stages at the transmitter; (i) $L_{\rm{us}}$-fold expansion of the delay-Doppler domain data symbols along the delay dimension, (ii) pulse-shaping the resulting signal along the delay dimension, (iii) IDFT operation across the Doppler dimension, and (iv) parallel-to-serial conversion by overlap-and-add operation along the delay dimension, i.e., overlapping the delay blocks every $M'$ samples. The reverse procedure is performed at the receiver.

%% file: BER_Results.tex
\section{Numerical Results}\label{sec:Numerical}
In this section, we numerically analyze and compare the OOB emissions and BER performance of the pulse-shaping techniques under investigation.
Additionally, we evaluate the efficacy of our proposed OOB reduction technique for C-PS OTFS that also improves the BER performance of all the pulse-shaping schemes. The observed staircase behavior of the ODDM spectrum confirms the validity of our derivations in Section~\ref{sec:LPS}.
In our simulations, we consider a delay-Doppler grid with $M=64$ delay bins and $N=32$ Doppler bins for data transmission at the carrier frequency $f_{\rm{c}}=5.9$~GHz, and the subcarrier spacing $\Delta f=15$~kHz. 
We employ a RRC pulse-shaping filter with the roll-off factor $0.1$ and the truncation parameter $Q=8$. The upsampling factor is set to $L_{\rm{us}}=2$.
We use the extended vehicular A (EVA) channel model \cite{3gpp}. 
The CP is chosen to be longer than the channel delay spread.
Maximum Doppler spread at the relative velocity of $v=500$~km/h between the transmitter and the receiver is considered, i.e., $\nu_{\rm{max}}=\frac{v}{c}f_{\rm{c}}$, where $c$ is the speed of light.
We assume perfect synchronization and perfect knowledge of the channel at the receiver. For detection, we deploy the minimum mean square error (MMSE) equalizer. 

\begin{figure}[!t]
  \centering 
  {\includegraphics[scale=0.29]{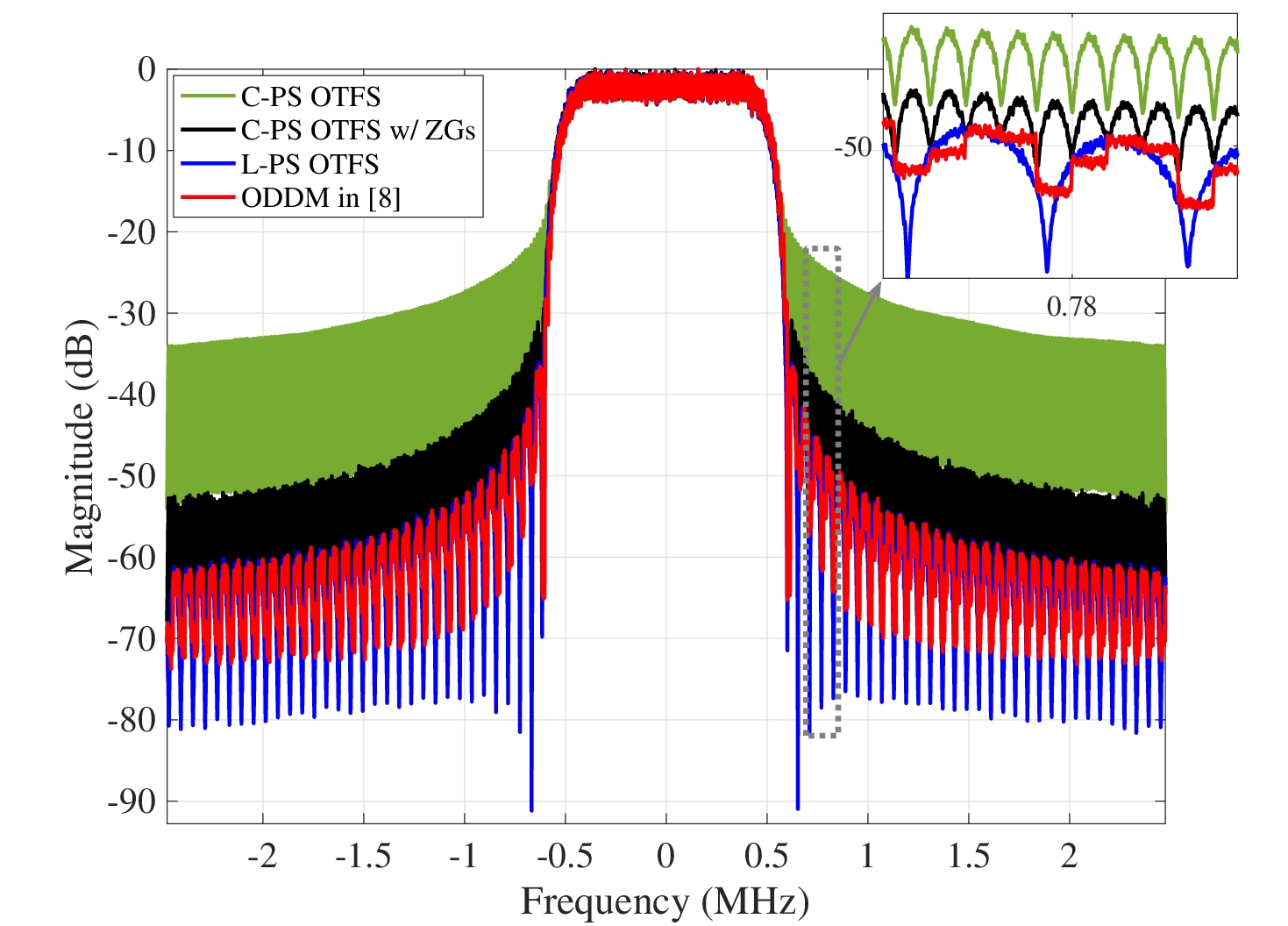}}
  \vspace{-0.3cm}
  \caption{PSD comparison of different pulse-shaping techniques for $M=64$, $N=32$, and $Q=8$.}
  \label{fig:PSD}
  \vspace{-0.5cm}
\end{figure}

In Fig.~\ref{fig:PSD}, we compare the OOB emissions of different pulse-shaping techniques. Fig.~\ref{fig:PSD} shows that the linear pulse-shaping techniques have the lowest OOB emissions. 
An interesting observation here is the staircase spectral behavior of ODDM which has not been previously reported in the literature. Based on our derivations, the modulated pulse in (\ref{eqn:Disc_DT2}) clearly explains this behavior. 
According to Section~\ref{sec:CPS}, C-PS~OTFS has large OOB emissions which is effectively reduced by ZG insertion along the delay dimension. 
As Fig.~\ref{fig:PSD} illustrates, by sacrificing only around $6\%$ of the delay block, i.e., $2$~ZGs at the edges of each delay block, up to $20$~dB reduction in OOB emissions is achieved. 
ZG insertion at the edges of each delay block is similar to full-CP OTFS where CPs (as guards) are inserted at the beginning of the delay blocks. In 5G~NR, the CP length is $7\%$ of the OFDM symbol duration. As a delay block in OTFS is equivalent to an OFDM symbol, allocating $6\%$ of the delay block length to the ZGs is inline with 5G~NR specifications.

\begin{figure}[!t]
  \centering 
  \vspace{0cm}
{\includegraphics[scale=0.29]{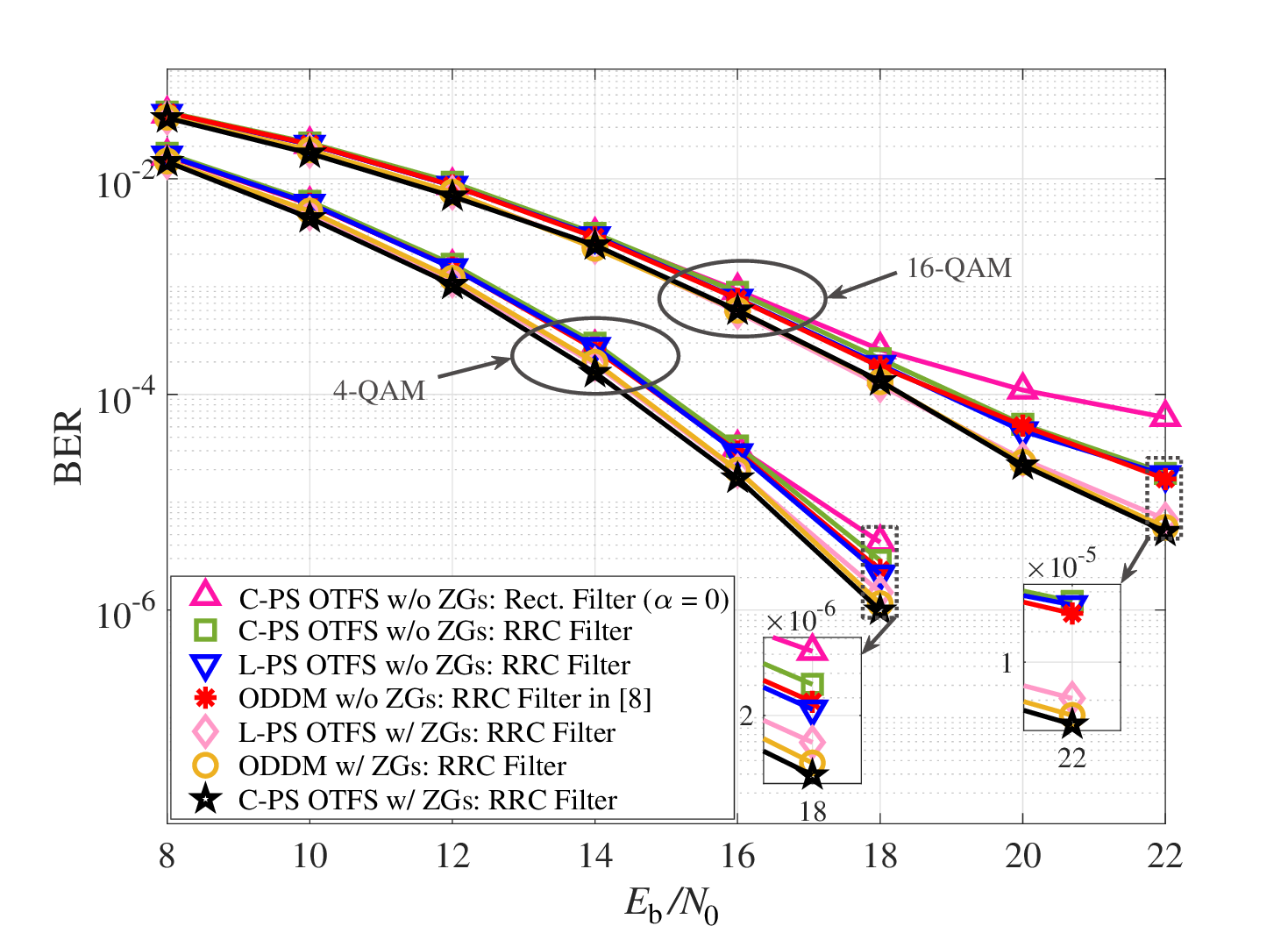}}
  \vspace{-0.3cm}
  \caption{BER performance of different pulse-shaping techniques for $M=64$, $N=32$, and $Q=8$ at the relative velocity of $v=500$~km/h.}
  \vspace{-0.3cm}
  \label{fig:BER_v500}
\end{figure}

In Fig.~\ref{fig:BER_v500}, we analyze the BER performance of different pulse-shaping techniques versus $E_{\rm{b}}/N_0$.
From Fig.~\ref{fig:BER_v500}, it is evident that both circular and linear pulse-shaping techniques yield nearly the same performance. 
As Fig.~\ref{fig:BER_v500} demonstrates, inserting only $2$ ZGs at the edges of each delay block effectively enhances the performance of all the pulse-shaping techniques by around $2$~dB for 16-QAM and $1$~dB for 4-QAM. For C-PS~OTFS, the rectangular pulse has the worst performance which is due to its large sidelobes.

%% file: Conclusion.tex
\vspace{-0.1cm}
\section{conclusion}\label{sec:Conclusion}
In this paper, we classified pulse-shaping of the multiplexed data on the delay-Doppler plane into two types of circular and linear pulse-shaping techniques. 
Then, we formulated each pulse-shaping  class by considering block-wise pulse-shaping along the delay dimension. 
We also formulated the recently emerged waveform ODDM in discrete time. Our derivations led to a generalized framework that explains all the pulse-shaping techniques under the same umbrella while revealing their properties, similarities and differences. 
Using our generalized framework, we clearly explained the reasons for high OOB emissions of circularly pulse-shaped signals. Then, we proposed an effective OOB emission reduction technique by inserting a small number of ZG symbols along the delay dimension. Furthermore, we showed that our proposed ZG insertion technique is also beneficial to the linear pulse-shaping techniques as it effectively improves their BER performance. 
We presented ODDM in our generalized framework which revealed its interesting staircase spectral behavior. We also showed that ODDM is a linear pulse-shaping technique.
Ultimately, we compared the OOB emission and BER performance of the pulse-shaping techniques by simulations while confirming our mathematical derivations and claims.